\documentstyle[epsf]{article}
\begin{document}
\baselineskip8mm
\title{\vspace{-6cm}Chaos in closed isotropic cosmological models
with steep scalar field potential}

\author{A. V. Toporensky}
\date{}
\maketitle
\hspace{-6mm}{\em Sternberg
Astronomical Institute, Moscow University, Moscow, 119899, Russia}\\
\begin{abstract}
The dynamics of closed scalar field FRW cosmological models
is studied for several types of exponentially and more than
exponentially steep potentials. The parameters of scalar field
potentials which allow a chaotic behaviour are found from numerical
investigations. It is argued that analytical studies of equation of
motion at the Euclidean boundary can provide an important information
about the properties of chaotic dynamics. Several types of transition
from chaotic to regular dynamics are described.
\end{abstract}
\\

\section{Introduction}
The studies of chaotical dynamics of closed isotropic cosmological
model has a long story. They were initiated by papers \cite{Ful-Park-Star}
where the possibility to avoid a singularity at the contraction stage
in such a model
with a minimally coupled massive scalar field
was discovered. Later it was found that this model allows
the existence of periodical trajectories \cite{Hawking} and aperiodical
infinitely bouncing trajectories having a fractal nature \cite{Page}.
 This result was reproduced in other terms in our
papers \cite{we,we1}.
 In \cite{Cornish} the
 set of  periodical trajectories was studied from the viewpoint of
 dynamical chaos theory. It was proved that the dynamics of a closed
universe with a massive scalar field is chaotic and an important
 invariant of the chaos, the topological entropy, was calculated.
 On the other hand, in our paper \cite{we1} we found
that deformations of the scalar field potential may change the
dynamical regime from chaotic to regular one. This fact poses a
question: is the chaotic behaviour closely connected with a concrete form
of the potential ($V(\varphi)=\frac{m^2 \varphi^2}{2}$) used in the
previous papers or is it a more general phenomenon?

A numerical analysis shows
that the situation for another natural scalar field potential
$V(\varphi)=\lambda \varphi^4$ is qualitatively the same. But most of modern
scenarios based on ideas of the string theory and compactification naturally
lead to another forms of potential which are exponential or behave as
exponential for large $\varphi$ (see, for
example, Ref. \cite{Zhuk}). This steepness of
the potential apparently changes the possibilities of escaping the
singularities and alters the structure of infinitely bouncing
trajectories.  Under some conditions, which will be studied below in
detail, the chaotic behaviour can completely disappear. Another
interesting problem concerning asymptotic regime near the singularity for
such potentials was recently studied in \cite{Foster}.

The structure of the paper is the following: in Sec. 2 we
recall the structure of chaos in the case of a massive scalar field
with a special attention paid to the statements which are valid for an
arbitrary scalar field potential. In Sec. 3 the conditions for
the chaotic behaviour in models with exponential potentials are
investigated both analytically and numerically. In Sec. 4 this
analysis is extended to steeper potentials. Sec. 5 provides a brief
summary of the result obtained.

 \section{Chaotic properties of closed FRW model with a scalar field}
We shall consider a cosmological model with an action
\begin{equation}
S = \int d^{4} x \sqrt{-g}\left\{\frac{m_{P}^{2}}{16\pi} R +
\frac{1}{2} g^{\mu\nu}\partial_{\mu}\varphi \partial_{\nu}\varphi
-V(\varphi)\right\}.
\end{equation}
For a closed Friedmann model with the metric
\begin{equation}
ds^{2} = N^{2}(t) dt^{2} - a^{2}(t) d^{2} \Omega^{(3)},
\end{equation}
where
$a(t)$ is a cosmological radius, $N$ -- a lapse function and
$d^{2} \Omega^{(3)}$ is the metric of a unit 3-sphere and
with homogeneous scalar field $\varphi$
the action (1) takes the form
\begin{equation}
S = 2 \pi^{2} \int dt N a^{3}
\left\{\frac{3m_{P}^{2}}{8\pi}
\left[-\left(\frac{\dot{a}}{N a}\right)^{2} + \frac{1}{a^{2}}\right]
+\frac{\dot{\varphi}^{2}}{2 N^{2}} - V(\varphi)\right\}.
\end{equation}

Now choosing the gauge $N = 1$ we can get the following equations of
motion
\begin{equation}
\frac{m_{P}^{2}}{16 \pi}\left(\ddot{a} + \frac{\dot{a}^{2}}{2 a}
+ \frac{1}{2 a} \right)
+\frac{a \dot{\varphi}^{2}}{8}
-\frac{a V(\varphi)}{4} = 0,
\end{equation}
\begin{equation}
\ddot{\varphi} + \frac{3 \dot{\varphi} \dot{a}}{a}
+ V'(\varphi) = 0.
\end{equation}
In addition, we can write down the first integral of motion of our system
\begin{equation}
-\frac{3}{8 \pi} m_{P}^{2} (\dot{a}^{2} + 1)
+\frac{a^{2}}{2}\left(\dot{\varphi}^{2} + 2 V(\varphi)\right)  =
0.
\end{equation}

It is easy to see from Eq. (6) that the points of maximal expansion
and those of minimal contraction, i.e. the points, where $\dot{a} =
0$ can exist only in the region where
\begin{equation}
a^{2} \leq \frac{3} {8 \pi}  \frac{m_{P}^2}{V(\varphi)} ,
\end{equation}
Sometimes, the region defined by inequalities (7) is called Euclidean
, and the opposite region is called
Lorentzian. This definition is not good enough, because for this
dynamical system there are
no classically forbidden regions at all
(see \cite{our} for detail), but we shall use it for
brevity. Now we would like to distinguish between the points of minimal
contraction where $\dot{a} = 0,$  $\ddot{a} > 0$ and those of maximal
expansion where $\dot{a} = 0,$   $\ddot{a} < 0$. Assuming
$\dot{a} = 0$, one can express $\dot{\varphi}^{2}$ from
Eq.  (6) as
\begin{equation}
\dot{\varphi}^{2} = \frac{3}{4 \pi} \frac{m_{P}^{2}}{a^{2}}
-2 V(\varphi).
\end{equation}
Substituting (8) and $\dot{a} = 0$ into (4) we have
\begin{equation}
\ddot{a} = \frac{8 \pi V(\varphi) a}{m_{P}^{2}}
-\frac{2}{a}.
\end{equation}
From Eq. (9) one can easily see that the possible points of
maximal expansion are localized inside the region
\begin{equation}
a^{2} \leq \frac{1}{4 \pi} \frac{m_{P}^{2}}{V(\varphi)}
\end{equation}
while the possible points of minimal contraction lie outside this
region (10) being at the same time inside the Euclidean region
(7).

 It is easy  to see that the value of $a$ on this separating curve
is $\sqrt{2/3}$ times smaller than the corresponding
$a$ on the Euclidean boundary for a given value of the scalar
field independently on the concrete form of $V(\varphi)$.  In
 \cite{our} it was shown that this ratio remains constant for
some cases
 in a more
general situation with a non-minimally coupled scalar field.

Here we would like
to describe briefly the approach presented in \cite{we}. The main idea
consists in the fact that in the closed isotropical model with a
minimally coupled scalar field satisfying the energodominance condition
all the trajectories have the
point of maximal expansion.  Then
the trajectories can be classified according to localization of their
points of maximal expansion. The points of maximal expansion
are all located inside the Euclidean region.
A numerical investigation shows that this area has a quasi-
periodical structure.
Narrow zones starting from which the trajectory
has the point of bounce are separated by wide zones
containing the initial conditions of trajectories
falling into a singularity.
Each zone of bounces
contains a periodical trajectory with the point of full stop ($\dot
a=0;$  $\dot \varphi =0$) on the Euclidean boundary.  Then studying the
substructure of these zones from the point of view of possibility to
have two bounces one can see that this substructure reproduce on the
qualitative level the structure of the whole region of possible points
of maximal expansion.  Continuing this procedure {\it ad infinitum}
yields the fractal set of infinitely bouncing trajectories. This kind
of fractal, nonattracting invariant set is typical for chaotic systems
without dissipation (see, for example, \cite{disc,Levin} ). In the
theory of dynamical chaos it is called strange repellor.

Numerical investigations show also
that the structure of periodical trajectories for the system Eqs (4)-
(6) have two important properties:

1. All the simple periodical trajectories (i.e. having only one bounce
per period) have a full stop point on the Euclidean boundary.

This property allows us to use the points of the Euclidean boundary and
zero velocities as an initial conditions for searching a possible
strange repellors.

2. Trajectories, going from the boundary into the euclidean region
has a point of maximal expansion almost immediately, and then go towards
a singularity. So, periodical trajectory approaches their bounce point
on the boundary from the Lorentzian side.

The only exception is a single peculiar periodical trajectory existing
in the case of nonzero cosmological constant term (see below).

This two properties were found in numerical analysis and the question
whether or not they are satisfied for an arbitrary scalar field
potential requires more investigation. Our researches show that it is
true at least for potentials steep as power-law and steeper.

If the aforesaid is satisfied, some analytical approach is possible.
Indeed,
periodical trajectories with a full stop point on the Euclidean
boundary penetrate into the Lorentzian region. That is why all such full
stop points
lie to the left from a critical point
introduced by Page \cite{Page} in the configuration space $(a,
\varphi)$ .  It is a point on the Euclidean boundary separating
trajectories going into Lorentzian and Euclidean regions.  By definition,
the Page's point is the point where the direction of motion at the
initial moment coincides with the direction of the tangent to the curve
given by equality in (7) , i.e.  the point where
 \begin{equation}
\frac{\ddot{\varphi}}{\ddot{a}} = \frac{d \varphi}{da}.
  \end{equation}
Using Eqs. (4), (5),and (7) one can find for the case
$V(\varphi)=\frac{m^2 \varphi^2}{2}$
 \begin{eqnarray}
&& \varphi_{page} = \sqrt{\frac{3}{4\pi}} m_{P}; \nonumber\\
&&a_{page}=1/m ,
\end{eqnarray}
 except for the
trivial solution $\varphi=0$, $a= \infty$.

Trajectories starting from $\varphi < \varphi_{page}$ and going
into the Euclidean region reach the point of maximal expansion almost
immediately after crossing the separating curve (see \cite{we} where we
studied such set of maximal expansion points). After that point the
trajectory goes towards singularity. So it looks like a small zigzag but
not a "true" bounce. To define a really significant bounce, Cornish and
Shellard in \cite{Cornish} used a condition that in the point of bounce
$a < 1/m$ in addition to $\ddot a >0;$   $\dot a=0$. This is just the
condition that bounces lie to the left from the Page point in
configuration space $(a, \varphi)$.

It is also possible to use the criterion that the trajectory
must return to the Lorentzian region  after the bounce
. Qualitatively the picture of regions
containing bouncing trajectories under this criterion is the same as
for the former one, with the only exception that
the width of the regions is somewhat smaller.
This criterion can be treated as more direct one, but it requires numerical
integration while the study of Page's points may be done analytically.
Indeed, it is easy to obtain an analog to (12) for an arbitrary potential
during the same procedures. The result is the following equation for the
$\varphi$-coordinate:
  \begin{equation}
 V(\varphi_{page})=\sqrt{\frac{3
m_{P}^2}{16 \pi}}V'(\varphi_{page}).
 \end{equation}

It can be easily derived from Eq.(13) that a qualitative picture of
massive scalar field case do not change for any pow-law potential with
the even index:  there is one non-trivial Page's point with the full
stop points of periodical trajectories lying to the left from it on the
euclidean boundary. In the next section we will see that the situation
changes significantly  for the exponentially steep potentials.

We finish this section by the descriptions of another important
modification of the scalar field potential -- introducing a constant
term (so called cosmological constant).  A non-zero value of the
cosmological constant $\Lambda$ makes possible an existence of
trajectories without points of maximal expansion, i.e. the trajectories
which begin and end in DeSitter regime
\begin{eqnarray}
 &&a(t) \sim \exp(-Ht),\;\;t \rightarrow -\infty;\nonumber \\
 &&a(t) \sim \exp(Ht),\;\;t \rightarrow \infty
\end{eqnarray}
or the trajectories
which begin in the singularity and end in DeSitter expansion or vice
versa. Here, in Eq. (14) $H$ denotes Hubble constant \[H =
\sqrt{\frac{\Lambda}{3}}.\]

The chaos in the model with a massive scalar field and a cosmological constant
\begin{equation}
V(\varphi)=\frac{m^2 \varphi^2}{2} + \frac{m_{P}^2}{8 \pi} \Lambda
\end{equation}
 was studied in \cite{we1}.  Before recalling some results of this
paper, let us look on Page's points for this potential.
The equation (13) has the form
\begin{equation}
\varphi^2-\sqrt{\frac{3 m_P^2}{4 \pi}} \varphi+\frac{\Lambda}{m^2}
\frac{m_{P}^2}{4 \pi}=0.
\end{equation}
 It is easy to see from (16) that for
$\Lambda/m^2>0.75$ there are no such points:  all trajectories from
the Euclidean boundary go into the Lorentzian region (and then fall into DeSitter
regime).  For $\Lambda/m^2<0.75$ there are two Page's points. The left point
has the same properties as for the potential without $\Lambda$, while
properties of the right point are inverse: the trajectories with an initial
$a>a_{page2}$ go into the Lorentzian region while trajectories with
$a<a_{page2}$ (but, of course,$a>a_{page1}$) -- to the Euclidean one.
A numerical analysis confirms that all trajectories with $a>a_{page2}$
fall to DeSitter regime and can not belong to the chaotic repellor. So, the
role of the 2-nd Page's point appear to be in restricting of the chaotic area in
the configuration space while the existence of chaos seems to be connected
with the 1-st one.

An unstable periodical trajectory,
found in \cite{we1} via numerical studies
that restricts the area of bounce intervals in $a$ from the large
values of the scale factor lies to the left from the 2-nd Page's point.
It crosses the $\varphi$-axis at some point with a scale factor $a_m$.

In addition to discussion in \cite{we1}, we give here a simple analytical
estimation of the location of this periodical trajectory in the configuration
space.  The typical value of $a_m$, corresponding to this trajectory
can be easily estimated for large $a$. In this case we may
assume the harmonic oscillations to be a solution for the field $\varphi$
dynamics:  $$ \varphi = A \cos (m t). $$ For calculating $A$ consider
a point where $\varphi=0; \dot a=0$.  Taking the derivative with
respect to time and using the constraint equation (6), we get

$$
A^2=\frac{m_{P}^2}{4 \pi} \frac{3/a^2-\Lambda}{m^2}.
$$

Substituting these two expressions into (4) and integrating
over one period $T$ of the scalar field oscillation, we
can find the velocity $\dot{a}$ after one oscillation:

$$
\dot{a}= T \frac{1}{2} (\Lambda a - \frac{1}{a}).
$$

It vanishes for
\begin{equation}
a_m=\sqrt{1/\Lambda}.
\end{equation}
 Trajectories with $a^2>1/\Lambda$ have
increasing scale factors during the scalar field oscillations.
They inevitably reach the DeSitter regime and can not belong to the
strange repellor.

As we found in \cite{we1},
if the cosmological constant is more than about one half
of the mass square of the scalar field (a more accurate value is
  $\Lambda > \sim 0.28 m^2$ ) the chaotic structure disappear and the
dynamics becomes regular. This critical ratio
$\Lambda/m^2$ is about $2.7$ times lower than that corresponding to the
disappearance of Page's points ($\Lambda/m^2=0.75$).

Surprisingly, the estimation (17) is correct even for
$\Lambda$ near the critical value when corresponding $a_m$ can
not be treated as large. Numerical values of $a_m$ differ from (17)
by several percents at most.

The possible values of $\Lambda$ allowing the chaotic dynamics for potential
(15) are also restricted from below for a negative cosmological constant.
The corresponding critical value is $|\Lambda|/m^2 \sim 0.34$ (see
\cite{our3} for detail).

\section {Chaotic motion in exponential potentials}
We start with one note about the pure exponential potential
$V(\varphi)=M_{0}^4 \exp(\varphi/\varphi_0)$. It is easy to see from Eq.(13) that
there are no Page's point in this case. The direction of a trajectory
starting with zero initial conditions from the Euclidean boundary is
fully determined by the value $\varphi_0$. If
 $\varphi_0 > \frac{\sqrt{3}}{4 \sqrt{\pi}} m_P$ {\it all
the trajectories} from the Euclidean boundary go into the Lorentzian region while
if $\varphi_0 < \frac{\sqrt{3}}{4 \sqrt{\pi}} m_P$
 all of them go into the Euclidean region.

Now we consider the potential  $V(\varphi)=M_{0}^4(\cosh(\varphi/\varphi_0)-1)$.
In the limit $\varphi \to 0$ it looks like $\frac{m_{eff}^2
\varphi^2}{2}$ with $m_{eff}=\frac{M_{0}^2}{\varphi_0}$, while for large
$\varphi$ it looks like the pure exponential one. The equation for
Page's point is now
\begin{equation}
\sinh{\frac{\varphi}{\varphi_0}}=\frac{\varphi_0}{\sqrt{3}}
\frac{4 \sqrt{\pi}}{m_P} (\cosh{\frac{\varphi}{\varphi_0}}-1).
\end{equation}

We can see that Page's point exists only if
$\varphi_0 > \frac{\sqrt{3}}{4 \sqrt{\pi}} m_P$. In the
opposite case, again all the trajectories from the Euclidean boundary go into
the Euclidean region. It means that there are no periodical trajectories
with the full stop points in this case.
Remembering the first property of our chaotic system listed in previous
section, we
may expect that the strange repellor
is absent at all when
$\varphi_0 < \frac{\sqrt{3}}{4 \sqrt{\pi}} m_P$.

\begin{figure}
\epsfxsize=\hsize
\centerline{\epsfbox{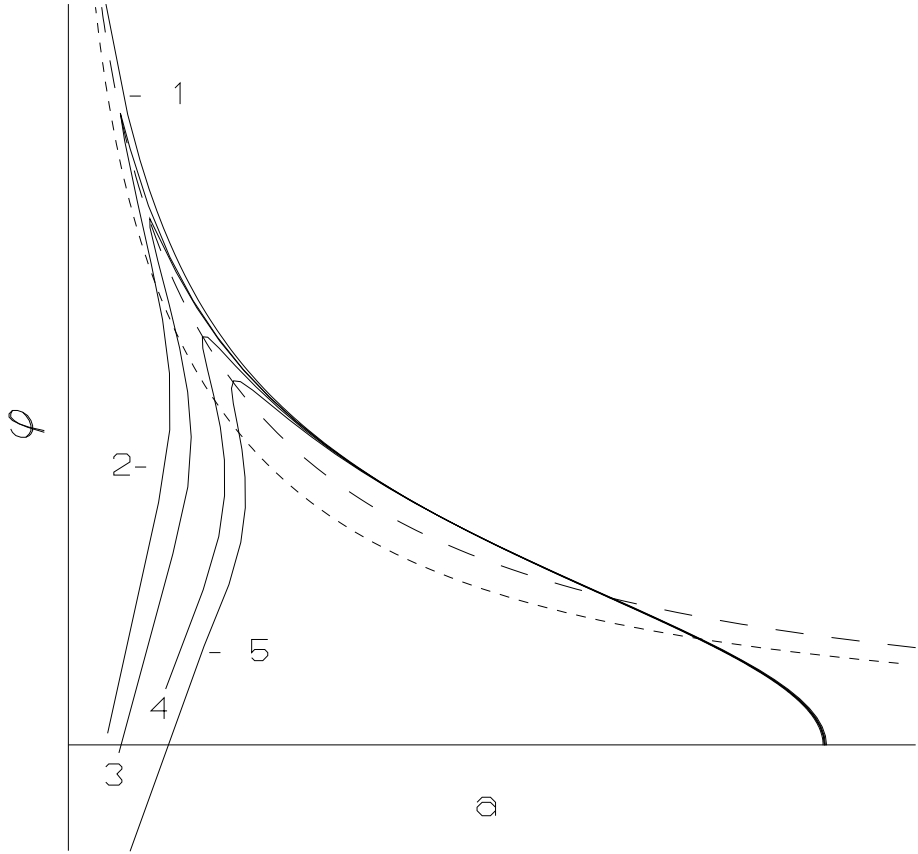}}
\caption{Example of trajectories with the initial conditions
close to
the boundary separating trajectories falling
into $\varphi=+ \infty$ (trajectory $1$) and $\varphi=-\infty$
(trajectories $2-5$) singularities for the case $\varphi_0 <
\frac{\sqrt{3}}{4 \sqrt{\pi}} m_P$. This boundary is sharp, no fractal
structure is present.  Trajectories $2-5$ have a zigzag-like form,
no periodical trajectories are present. The long-dashed line is the
Euclidean boundary, the short-dashed line is the separating curve.}
\end{figure}

Results of the numerical integration confirms  this analytical
considerations. In Fig.1 several typical trajectories are plotted. Although
points with $\dot a =0;$  $\ddot a >0$ are possible, trajectories
containing such points have a zigzag-like form
and can not return to  the Lorentzian
region. Periodical trajectories are absent and the dynamics, in contrast
to previously considered cases, is regular: it is impossible to avoid a
singularity even on zero-measure set of initial condition. All the
trajectories fall into singularity.

For $\varphi_0 > \frac{\sqrt{3}}{4 \sqrt{\pi}} m_P$
 the structure of trajectories is
similar to the massive scalar field case. Again, if we consider
points of maximal expansion, we can find regions
corresponding to the bouncing trajectories. To distinguish
between bounces and
zigzags, we use an additional condition that the value of $\varphi$ at
the point of bounce is greater than $\varphi_{page}$ (or we can
use the condition that the trajectory after the bounce returns
 to the Lorentzian
region). Then in limit
$\varphi_0 \to \frac{\sqrt{3}}{4 \sqrt{\pi}} m_P$ the width of
bouncing regions tends to zero. It should be noted that when
$\varphi_0$ crosses the critical value the entire structure of periodical
trajectories disappears in a jump-like manner.

 Now we turn to a more
general case $V(\varphi)=M_{0}^4(\cosh(\varphi/\varphi_{0})-\Lambda)$.
It corresponds to an effective
cosmological constant $\Lambda_{eff}=\frac{8 \pi M_{0}^4}{m_{P}^{2}}
(1-\Lambda)$.  Let us study the case of positive $\Lambda_{eff}$.
Page's points exist if the transcendent equation \begin{equation}
\sinh{\frac{\varphi}{\varphi_0}}=\frac{\varphi_0}{\sqrt{3}}
\frac{4 \sqrt{\pi}}{m_P} (\cosh{\frac{\varphi}{\varphi_0}}-\Lambda)
\end{equation}
have roots.

\begin{figure}
\epsfxsize=\hsize
\centerline{\epsfbox{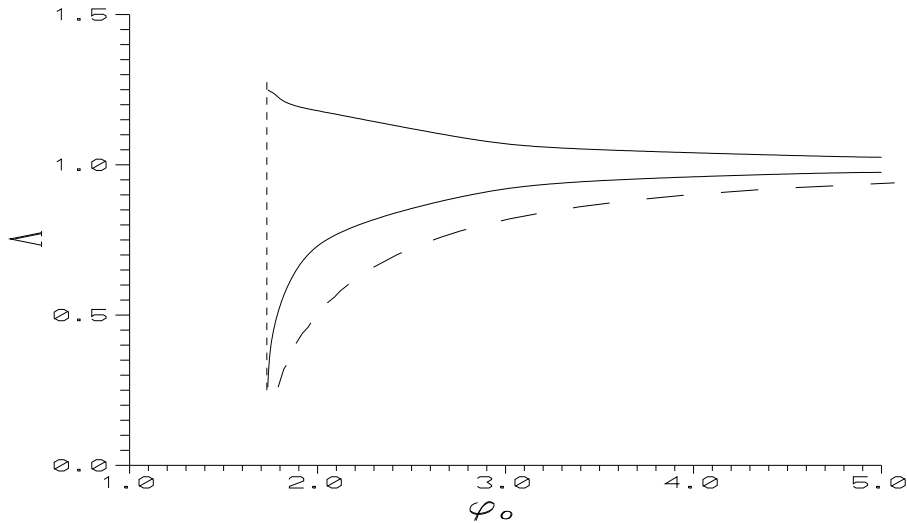}}
\caption{
Several regions in the parameter plane for two-parameter
family (19) which relate to a possibility of the chaotic dynamics.
The parameter $\varphi_0$ is measured in units $m_P/\sqrt{16 \pi}$.
For $\varphi_0 < \sqrt{3}$ (to the left from the short-dashed curve)
the dynamics is regular. For $\varphi_0 > \sqrt{3}$ the chaotic
dynamics exists for $\Lambda$ lying between two solid lines.  The
Page's points are absent below the long-dashed curve.}
\end{figure}

If $\varphi_0 < \frac{\sqrt{3}}{4 \sqrt{\pi}} m_P$
 this equation has one root. The Page's point corresponding to this sole
root looks like the 2-nd one for a massive scalar field with
a cosmological constant:  trajectories with $a>a_{page}$ tend to the
Lorentzian region while with $a<a_{page}$ -- to the Euclidean one. And
the dynamical picture obtained from the numerical integration have
similar features: one unstable periodical orbit near $a^2=1/\Lambda_{eff}$
and nothing else. The discrepancy between the estimation (17) and the
numerical solution grows with $\varphi_0$ reaching it's critical
value.

The case $\varphi_0 > \frac{\sqrt{3}}{4 \sqrt{\pi}} m_P$
is fully similar to the massive one with a cosmological constant:
there exist two or zero Page's points depending on the values of
$\varphi_0$ and $\Lambda$ (the long-dashed curve in Fig.2).
The chaos disappears
for $\Lambda$ (the lower solid curve in the same plot) which are
somewhat larger than
needed for Page's points to disappear.

For completeness we also present the area of values of
$\Lambda_{eff}<0$ leading to the chaotic dynamics (it is bounded
by the upper solid curve in Fig.2). It essentially depends
on the trajectory behaviour for large values of the scale factor and hence
can not be understood by studying Page's points. We see from this plot that
the massive case relation $|\Lambda_{eff}|/m_{eff}^{2} \sim 0.34$ is still valid
with a good accuracy.

\section{Very steep potentials}
For potentials more steep than exponential we face a situation which
differs from the massive scalar field case significantly. It can be shown
from studying the analog of (11) that
trajectories from the Euclidean boundary for large values of $\varphi$
go into Euclidean region. So the area of possible periodical
trajectories is limited from the side of large $\varphi$. This leads to limiting
the possible values of the scale factor on such trajectories. In Fig.3
we schematically show the number of oscillations of the field $\varphi$ for
periodical trajectories depending on the location of their
full stop points on the Euclidean boundary. In contrast to the previous cases, this
number is restricted from above by some number $N$.  The numerical results
show that the number of bounce intervals is also finite and restricted
by some number $M>N$. In addition, there exists a complicated system of
rules determining the substructure of intervals because the number of
subintervals is now depends on the ordinal number of the interval. In
particular, for an interval with ordinal number $N_1>N$, the number of
subintervals is always less than $N$.  It is interesting that this
dynamical picture looks very similar to that described in \cite{our3} for
the system with a massive scalar field and a hydrodynamical matter.

\begin{figure}
\epsfxsize=\hsize
\centerline{\epsfbox{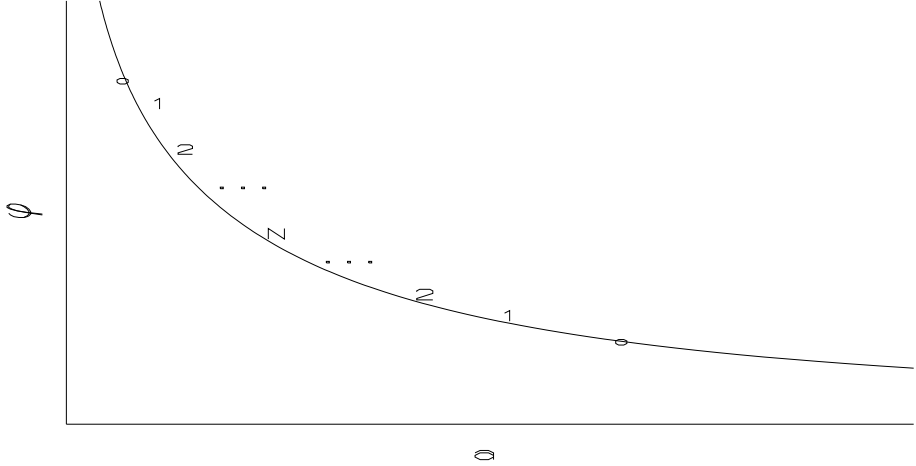}}
\caption{
The relation between the number of the scalar field
oscillations for trajectories starting from the Euclidean boundary and
the position of their initial point in the case of very steep scalar
field potentials.}
\end{figure}

Numerical investigations also show the transitions to regular dynamics
which can be understood by studying the Page points configuration.  Let us
illustrate this picture using some concrete one-parameter family of
potentials $$ V(\varphi)=
M_{0}^4 (\exp(\varphi^2/\varphi_{0}^2)+\exp(-\varphi^2/\varphi_{0}^2)-2).
$$
Depending on the parameter $\varphi_0$, two or zero
nontrivial Page,s points can appear. The critical value of $\varphi_0$ obtained
numerically is $\varphi_0 \sim 0.905 m_P$.  For $\varphi$ larger than this
value we have two Page's points. The area of initial points for
trajectories penetrating into the Lorentzian region (i.e. possible full
stop points of periodical trajectories) lies between them. So the properties
of the Page's points are inverse with respect to those for the case of
a massive scalar field with a cosmological constant.

In reality, the
chaotic behaviour disappears for a somewhat larger value $\varphi \sim 0.96 m_P$.
It is however remarkable that this quite simple analytical expressions
(with computer used only to solve corresponding transcendent equation)
gives us a rather good estimation of values of $\varphi_0$ which allows
a chaotic regime.

The concrete form of potential for large $\varphi$ does not affect  the
properties of chaos and therefore the steepness of the potential
can be arbitrary high. The picture described above is valid even
for cases with infinitely high potential walls. We have investigated a
simple case $$ V=A/(\varphi_{0}^2-\varphi^2)-A/\varphi_{0}^2. $$ In this case
the condition of the existence of Page's points has the exact solution
$\varphi_0>\frac{9}{4 \sqrt{\pi}} m_P$,
while the numerical result for the existence of chaotic dynamics is
$\varphi_0>\frac{9.6}{4 \sqrt{\pi}} m_P$.

\section{Conclusions}
  We have studied the dynamics of closed Friedmann-Robertson-Walker
universes with a scalar field.
It was found that there exists a rather wide class of scalar field
potentials, more steep than power-law, for which the dynamics of
a scale factor is regular and there is no possibility to escape a
singularity at a contraction stage.

On the other side, the class of potentials which allow
 the chaos,
associated with chaotic oscillations of the scale factor, is also
sufficiently wide.  This chaos manifests itself in the presence of a
strange repellor - a fractal set of unstable periodical orbits. Their
existence is connected with a possibility for this system to have a
bounce.  Taking the point of maximal expansion as the initial one, the
regions of the initial conditions in configuration space that lead to
bouncing trajectories have a rather regular and obvious structure: the
N-th region contains trajectories which have a bounce after N
oscillations of field $\varphi$.  More investigations is still
required to study the possibility of existence of more complicated
chaos for some another class of scalar field potentials.

The formal definition
of the bounce ($\dot a=0$, $\ddot a > 0$)
is not good enough. The reason is that trajectories like in Fig.1
is very natural for our dynamical system. They satisfy the formal
bounce criterion, but it is intuitively clear that zigzags on
trajectories shown in Fig.1 can not change significantly the fate of
the trajectories falling into singularity.  On the other hand, the
condition we used previously \cite{we} (the value of scale factor at
the next point of maximal expansion is greater than at previous one)
rejects trajectories with decreasing but lying inside the area of the
strange repellor value of $a$ at the points of maximal expansion.
The condition that the trajectory
after the bounce returns to the Lorentzian region or that the scale
factor $a$ at the point of bounce is less than the scale factor of the
chaos generating Page's point may be used as a more suitable additional
criterion.

Varying the potential, several types of transitions from chaotic
to regular behaviour can be distinguished:
(1) The number of bounce regions in the initial condition space can
remains infinite, but the area of the strange repellor in
the configuration space tends to zero (for the potential like
$V(\varphi)=\frac{m^2 \varphi^2}{2} + \Lambda$; $\Lambda>0$). (2) The
number of the bounce regions becomes finite and tends to zero while
the area of the strange repellor remains unbounded (the same
potential but for $\Lambda<0$). (3) The number of the bounce regions
diminishes and the strange repellor shrinks (for very steep potentials
or for a mass-like potential in a more general case of a scalar field
with a hydrodynamical matter \cite{our3}) (4) The width of the bounce
regions tends to zero while their number and the area of the
strange repellor remain unchanged (for
$V(\varphi)=M_{0}^4 (\cosh(\varphi/\varphi_0)-1)$).  The study of the
transition from chaos to order for such dynamical systems
may be interesting from the mathematical
point of view.

Using the fact that the region in the configuration space where bounces
are possible represents a sufficiently narrow area near the
Euclidean boundary, we have proposed that studying trajectories having
full stop points at this curve can provide us with an important
information about the strange repellor as a whole. This suggestion
realizes at least for the type of chaos which satisfies two properties,
established by numerical investigations for steep potentials and
described in Sec.2. Their validity for less steep
potentials will be the goal of our future work.
In addition, the
transition from chaos to order appear to be closely connected with the
change of structure of Page's point configuration at the Euclidean
boundary at least for positive scalar field potentials.

\section*{Acknowledgement}
 This work was supported by Russian Foundation for Basic Research
 via grants No 96-02-16220 and No 96-02-17591.

\end{document}